\documentstyle[11pt,newpasp,twoside,psfig]{article}
\index{instructions}
\index{guidelines}

\def\edcomment#1{\iffalse\marginpar{\raggedright\sl#1\/}\else\relax\fi}
\reversemarginpar

\begin{document}
\title{Long-Term Timing of 374 Pulsars}
 \author{G. Hobbs}
\affil{Australia Telescope National Facility -- CSIRO, P.O. Box 76,
 Epping NSW 1710, Australia}
\author{A. Lyne, M. Kramer}
\affil{Jodrell Bank Observatory, University of Manchester,
 Macclesfield, Cheshire, SK11 9DL, UK}

\begin{abstract}
We have analysed the arrival times for 374 pulsars that have been
observed for more than six years using the 76-m Lovell telescope at 
Jodrell Bank Observatory.  Here we present a qualitative analysis of
structures seen in the timing residuals.
\end{abstract}

\section{Introduction}

More than 500 pulsars are being regularly observed using the 76-m
Lovell radio telescope at Jodrell Bank Observatory at frequencies
between 408 and 1630\,MHz.  This archive contains more than 5600 years
of pulsar rotational history which we supplement, for 18 pulsars, with
early observations from the Jet Propulsion Laboratory (Downs \&
Reichley 1983) to provide individual data spans of up to 34 years. In
a series of three papers we plan to carry out a full analysis of the
data for more than 350 pulsars that have been observed for longer than six
years.  The first paper provides accurate timing solutions including
proper motion measurements.  In the second paper we will use these
proper motion values to improve our understanding of pulsar
velocities.  In the third paper we will discuss the remnant structures
in the timing residuals after fitting a timing solution for rotational
frequency and its first derivative.  Some qualitative results, which
will be discussed in more detail in the third paper, are highlighted
here.

\section{Structure in the timing residuals}

Many different structures are observed in the timing residuals,
including cubic features, cusps, flat residuals, pseudo-sinusoids and
more irregular features.  Cusps (corresponding to a local maxima)
indicate the pulsar undergoing a glitch. Pseudo-sinusoidal
variations in the timing residuals may be caused by the pulsar
precessing or from unmodelled binary companions.  

The amplitudes of the observed cubic structures are many orders of
magnitude greater than those predicted from magnetic dipole radiation;
measured `braking indices' range between $\pm 3\times10^8$.  

The structures observed in the majority of the timing residuals are
not time resolved.  For example, about 30\% are dominated by a large
scale cubic feature and a further 40\% currently have no clearly
defined features.  However, we have a sample of 29 pulsars for which
it is possible to study individual features in detail.  We find that
the distribution of the residuals around the mean value has no general
asymmetry (with 15 having negative skew and 14 positive).  We do find
that 70\% have, on average, sharper (larger second derivative) local
maxima than local minima.  No explanation is given for this effect at
the present time.

The time corresponding to the first local minima in the
auto-correlation function of the timing residuals provides an estimate
of the time--scale of the structures present.  However, we see no
correlation with this parameter and the observed or derived pulsar
parameters.  We do find strong correlations between the amplitude of
the timing noise and the first and second rotational frequency
derivatives (Figure 1).

\begin{figure}
\begin{center}
\begin{tabular}{cc}
\psfig{file=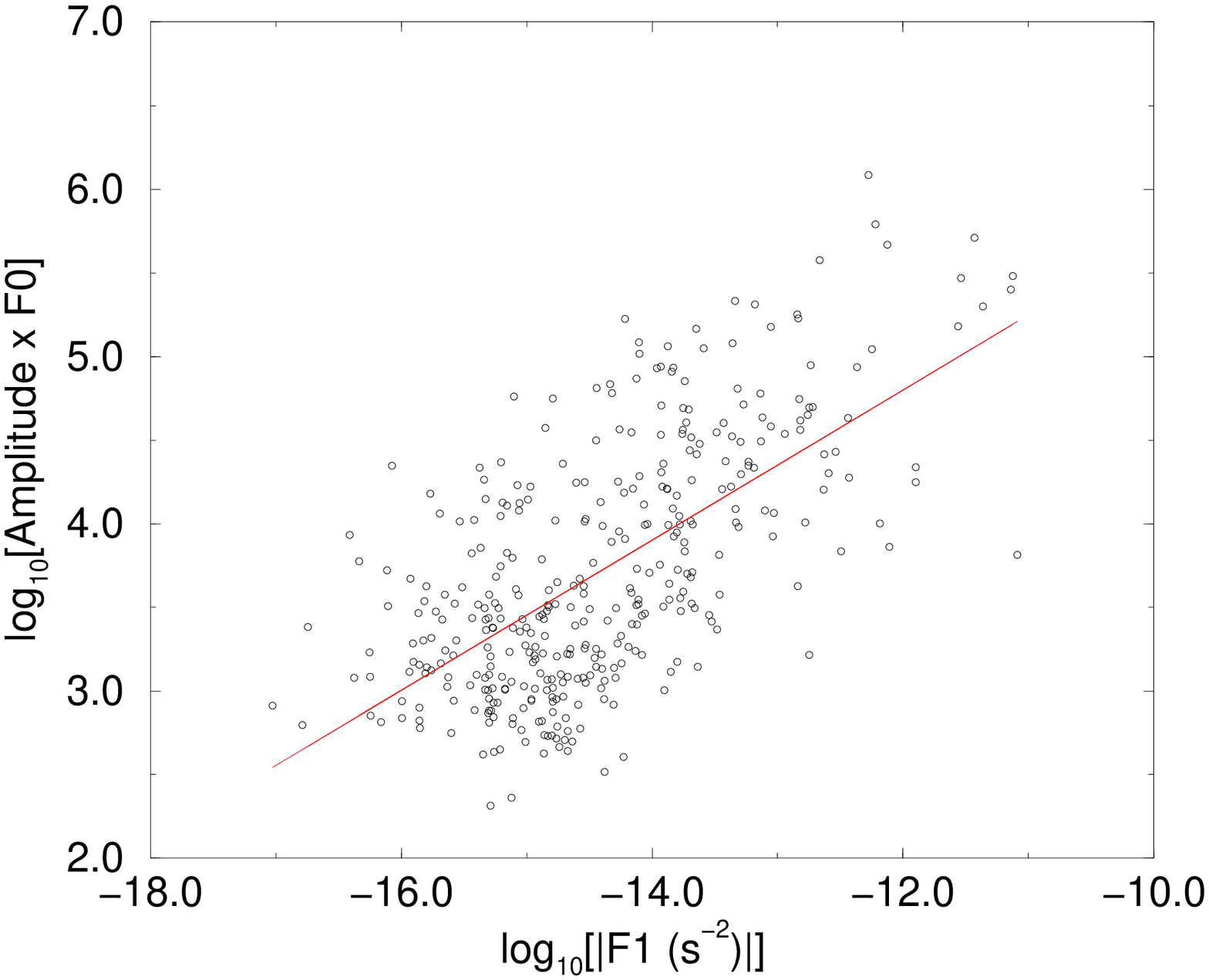,width=6.5cm,angle=-90} & 
\psfig{file=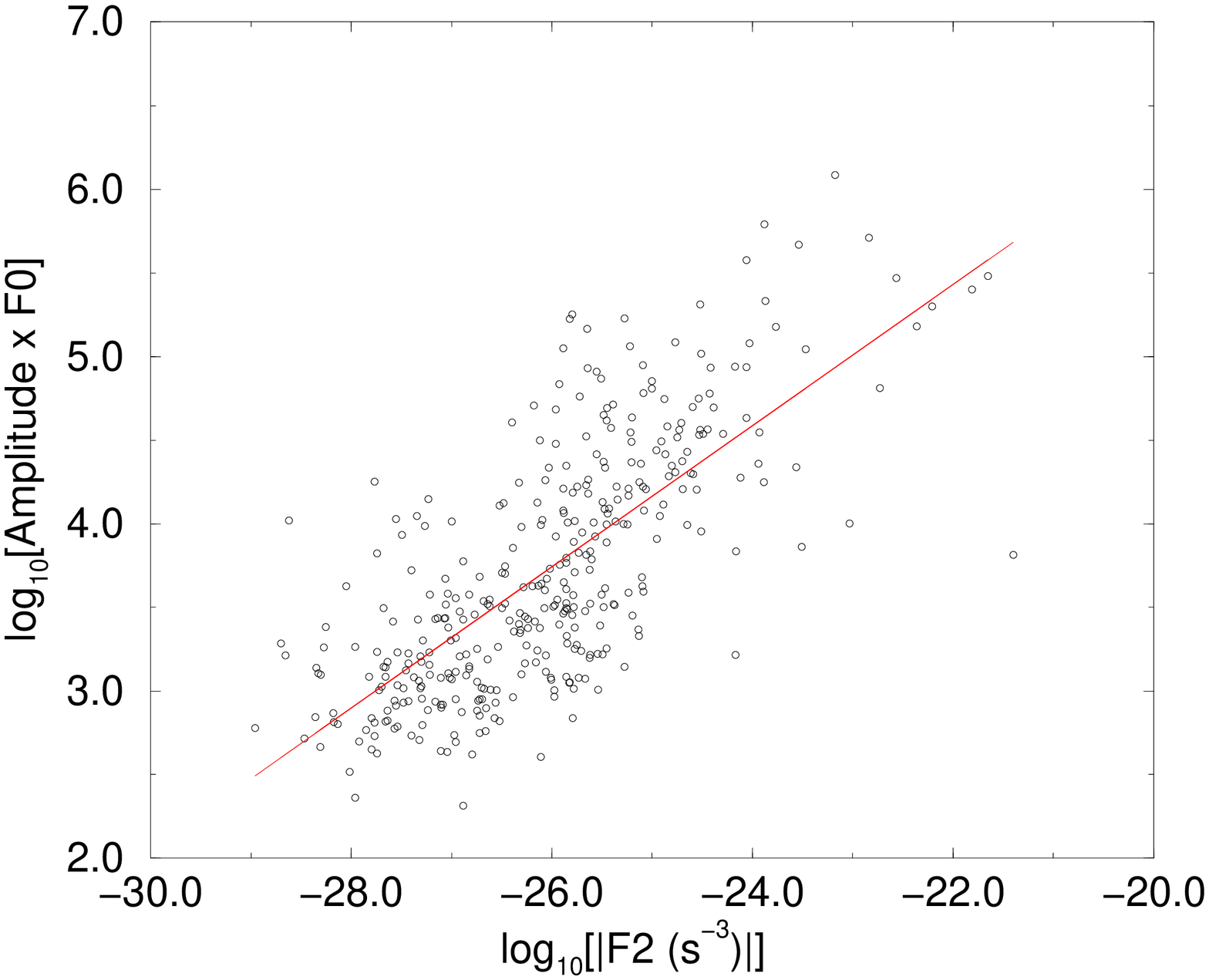,width=6.5cm,angle=-90} \\
\end{tabular}
\caption{The rms amplitude of the timing noise plotted against
rotational frequency first and second derivatives.}
\end{center}
\end{figure}

Our new sample of timing residuals that span up to $\sim$30\,years
will allow the first large scale analysis of timing noise to be
carried out.  Already it is clear that many of the structures observed
in the timing residuals can not have been caused by a simple `noise'
process. 

\section{Acknowledgements}

Many people have been involved in timing pulsars at Jodrell Bank.  We
thank, in particular, Christine Jordan for her work in developing much
of the software used in the analysis of the observations.

\end{document}